\documentclass[preprint,5p]{elsarticle}
\pdfoutput=1
\usepackage{natbib}
\usepackage{amsmath}
\usepackage{amssymb}
\usepackage{amsfonts}
\usepackage{hyperref}
\usepackage{epsfig}
\usepackage{graphicx}
\usepackage{slashed} 
\usepackage{mathrsfs} 
\usepackage{subcaption}
\usepackage{ulem}
\usepackage{color}

\def\prd{Phys. Rev. D}
\def\mnras{MNRAS}
\def\apj{ApJ}
\def\apjl{ApJL}

\def\aap{A\&A}

\def\jcap{JCAP}
\def\pasa{PASA}               

\begin{document}

\begin{frontmatter}

\title{Dark matter annihilation in $\omega$ Centauri: astrophysical implications
   derived from the MWA radio data}

\author[label1]{Arpan Kar\corref{mycorrespondingauthor}}
\cortext[mycorrespondingauthor]{Corresponding author}
\ead{arpankar@hri.res.in}
\author[label2]{Biswarup Mukhopadhyaya}
\author[label3]{Steven Tingay}
\author[label3,label4]{Ben McKinley}
\author[label5]{Marijke Haverkorn}
\author[label3]{Sam McSweeney}
\author[label3]{Natasha Hurley-Walker}
\author[label6]{Sourav Mitra}
\author[label7]{Tirthankar Roy Choudhury}

\address[label1]{Regional Centre for Accelerator-based Particle Physics, Harish-Chandra Research Institute, HBNI,
Chhatnag Road, Jhunsi, Allahabad - 211 019, India}

\address[label2]{Indian Institute of Science Education and Research Kolkata, Mohanpur, West Bengal 741246, India}

\address[label3]{International Centre for Radio Astronomy Research, Curtin University, Bentley, WA 6102, Australia}

\address[label4]{ARC Centre of Excellence for All Sky Astrophysics in 3 Dimensions (ASTRO 3D), Bentley, Australia}

\address[label5]{Department of Astrophysics/IMAPP, Radboud University, P.O. Box 9010, 6500 GL Nijmegen, The Netherlands}

\address[label6]{Surendranath College, 24/2 M. G. ROAD, Kolkata, West Bengal 700009, India}

\address[label7]{National Centre for Radio Astrophysics, TIFR, Post Bag 3, Ganeshkhind, Pune 411007, India}

\begin{abstract}
We present an analysis of Murchison Widefield Array radio telescope data from  $\omega$ Cen, possibly a stripped dwarf spheroidal  galaxy core captured by our Galaxy.  Recent interpretations of 
Fermi-LAT $\gamma$-ray data by  Brown {\it et al.} (2019) and Reynoso-Cordova {\it et al.} (2019) suggest
that $\omega$ Cen may contain significant Dark Matter.  We utilise  their best-fit Dark Matter annihilation models, and an estimate of the magnetic field strength in $\omega$ Cen, to calculate the expected radio synchrotron signal from annihilation, and  show that one can usefully rule out significant parts of the magnetic field - diffusion coefficient plane using our current observational limits on the radio emission. Improvement by a factor of 10-100 on these limits could constrain the models even more tightly.  
\end{abstract}


\end{frontmatter}

\section{Introduction}

$\omega$ Cen is possibly a stripped dwarf spheroidal galaxy core captured by our Galaxy, in which the dark matter (DM) density may be high. Two groups have recently 
analysed the $\gamma$-ray data from Fermi-LAT for this object, to suggest best-fit values
of the DM mass and annihilation rates in various dominant channels. We take advantage of the proximity  of
$\omega$ Cen and calculate the expected radio synchrotron surface brightness arising from DM annihilation in the ambient magnetic field,
corresponding to each of the fits. In each case, Murchison Widefield Array (MWA) data constrain the magnetic field - diffusion coefficient plane for $\omega$ Cen. 
It is of particular significance that some knowledge of the magnetic field is used here to constrain the diffusion coefficient, which is difficult to estimate.

Previously we have reported the results of the first low radio frequency search for the synchrotron emission signal expected from some DM annihilation models \citep{2019PhRvD.100d3002K}.  
Such radio signals arise from electrons and positrons, produced in DM annihilation cascades, undergoing cycloidal motion in the ambient magnetic field.
We targeted 14 DM-rich dwarf spheroidal (dSph) galaxies using observations from the 
MWA \citep{2013PASA...30....7T} and the Giant Metre-wave Radio Telescope (GMRT \citep{2017A&A...598A..78I}) to place limits on  diffuse synchrotron emission from these galaxies
for various models. While this represented a significant step toward much deeper observations by
the future Square Kilometre Array (SKA) \cite{Kar:2019cqo, Cembranos:2019noa, PhysRevD.99.021302, Colafrancesco:2015ola}, the results placed only limited constraints on DM models.  Similar results have recently been reported by a team using another low frequency interferometer, LOFAR, for the dSph galaxy Canes Venatici I \citep{2019arXiv190912355V}. One fundamental factor affecting our previous results and the LOFAR results is, of course, the large distances to dSph galaxies.

\section{Observations of Omega Centauri ($\omega$ Cen)}

A case has been recently made by Ref. \cite{2019arXiv190708564B} and Ref. \cite{Reynoso-Cordova:2019biv} that $\omega$ Cen, historically classified as the largest globular cluster associated with our Galaxy, is the captured and stripped core of a dSph galaxy and has a significant DM component to its mass. These characteristics potentially make 
$\omega$ Cen a suitable object for studying DM annihilation models, since it is only 5.4 kpc from the Earth.

Ref. \cite{2019arXiv190708564B} and Ref. \cite{Reynoso-Cordova:2019biv} analyze the Fermi-LAT data on $\gamma$-ray emission from the direction of $\omega$ Cen and claim consistency of the signal with DM annihilation. 
Ref. \cite{2019arXiv190708564B} claims the best fit to be $m_{\rm DM}=31\pm4$ GeV and a velocity-averaged annihilation cross-section of
$\rm{log_{10}}[\langle \sigma v \rangle~(cm^{3}s^{-1})]=-28.2\pm^{0.6}_{1.2}$,
with $b{\bar b}$ as the principal annihilation channel. The fit by Ref. \cite{Reynoso-Cordova:2019biv}, 
on the other hand, favours $m_{\rm DM}=9.1\pm^{0.69}_{0.62}$ GeV and  $\rm{log_{10}}[\langle \sigma v \rangle J~(GeV^2cm^{-2}s^{-1})]=-5.5\pm0.03$ for the $q{\bar q}$ channel, or $m_{\rm DM}=4.3\pm^{0.09}_{0.08}$ GeV and $\rm{log_{10}}[\langle \sigma v \rangle J~(GeV^2cm^{-2}s^{-1})]=-4.34\pm0.03$) for a $\mu^+ \mu^-$ channel. 
These best-fit values assume that the $\gamma$-ray signal from $\omega$ Cen 
is arising solely due to DM annihilation. Since our 
emphasis is on the corresponding radio synchrotron signals, we have used
this at face value.

No obvious populations of conventional high energy astrophysical objects are known to be associated with $\omega$ Cen to readily explain the $\gamma$-ray emission, although a more comprehensive examination of possible high energy photon sources is clearly required.  For example, Ref. \cite{2018MNRAS.479.2834H} find 30 objects in $\omega$ Cen with X-ray luminosities and colors consistent with a millisecond pulsar interpretation and suggest that these objects could be the source of the $\gamma$-ray emission seen with Fermi-LAT.

In order to predict synchrotron emission due to DM annihilation, the ambient magnetic field strength is an important parameter. The determination of magnetic field strength in dSph galaxies is not straightforward. However, measurements of the magnetic field in globular clusters are possible through the observation of pulsars, in particular via measurement of their dispersion measure and rotation measure. For example, Ref. \cite{2016mks..confE..36P} find evidence for magnetic fields possibly as high as 200 $\mu G$ in the globular cluster 47 Tucanae.  While a similar analysis is not currently possible for $\omega$ Cen (no pulsars are known in $\omega$ Cen), approximate estimates of its magnetic field strength are possible by virtue of its location within our Galaxy.

Using the best-fit DM annihilation models, suggested by 
Ref. \cite{2019arXiv190708564B} and Ref. \cite{Reynoso-Cordova:2019biv} noted above (obtained by using the best-fit J-factor, as explained later in this paper), the proximity of $\omega$ Cen, and the prospect of independently estimating the magnetic field strength within $\omega$ Cen (as outlined below), we re-visit the techniques used in our previous work to compute the corresponding radio synchrotron annihilation signals at low frequencies. The reader is referred to Ref. \cite{2019PhRvD.100d3002K} for the details of our observational approach, a summary of the relevant literature, and our results for dSph galaxies.
Finally, we compare our radio data with the theoretical expectations.

As has been mentioned above, our emphasis is on the
correlation between gamma-ray and radio signals from a nearby globular
cluster like Omega Centauri. The new step that we take in this study is
the calculation of the radio flux for the same dark matter profile(s), 
based on which gamma-ray data have been interpreted,   
and  its comparison with the MWA data. Thus \cite{2019arXiv190708564B} and \cite{Reynoso-Cordova:2019biv}
serve as our reference models for 
gamma-ray emission from dark matter annihilation in
Omega Centauri. Going beyond them in a study like this would have
taken us  to territories
where the aforementioned correlation would be impossible. 
Thus we found
it sensible to confine ourselves to the modelling in [9] and [10].
We are thus able to relate the interpretation of gamma-ray data in each 
of these analyses to the corresponding radio data and their implication for the
astrophysical parameters in Omega Centauri. While models differing from 
those in Ref. \cite{2019arXiv190708564B} and Ref. \cite{Reynoso-Cordova:2019biv} can certainly be there, our approach can be used
to predict and analyse the concomitantly altered radio synchrotron flux
for each of them. Thus the adherence to particular models used in
the recent literature enables us to lay out a general principle.

\section{Radio data analysis}

Following Ref. \cite{2019PhRvD.100d3002K}, we  produce an image for $\omega$ Cen that represents 
only extended, diffuse radio emission at 200 MHz, via the difference between an MWA image (sensitive to diffuse emission) and an 
image from the GMRT TGSS ADR1 (sensitive to compact emission). $\omega$ Cen is located close to the powerful radio galaxy 
Centaurus A (Cen A) and the MWA GLEAM data utilised in Ref. \cite{2019PhRvD.100d3002K} 
cannot be utilised here, the reason being the difficulties of producing high quality images in the vicinity 
of extremely bright and complex objects such as Cen A. Thus, for $\omega$ Cen, we utilise MWA 
images that were produced specifically for Cen A, by \cite{2018MNRAS.474.4056M}.  
We note that the GMRT data for $\omega$ Cen are not affected by Cen A in the same way that the MWA data are, due to the fact that the GMRT field of view is far smaller than the MWA field of view, meaning that the structure of Cen A that challenges MWA imaging is greatly attenuated by the primary beam response of the GMRT.

\begin{figure}[t!]
\centering
  \includegraphics[height=0.3\textwidth, angle=0]{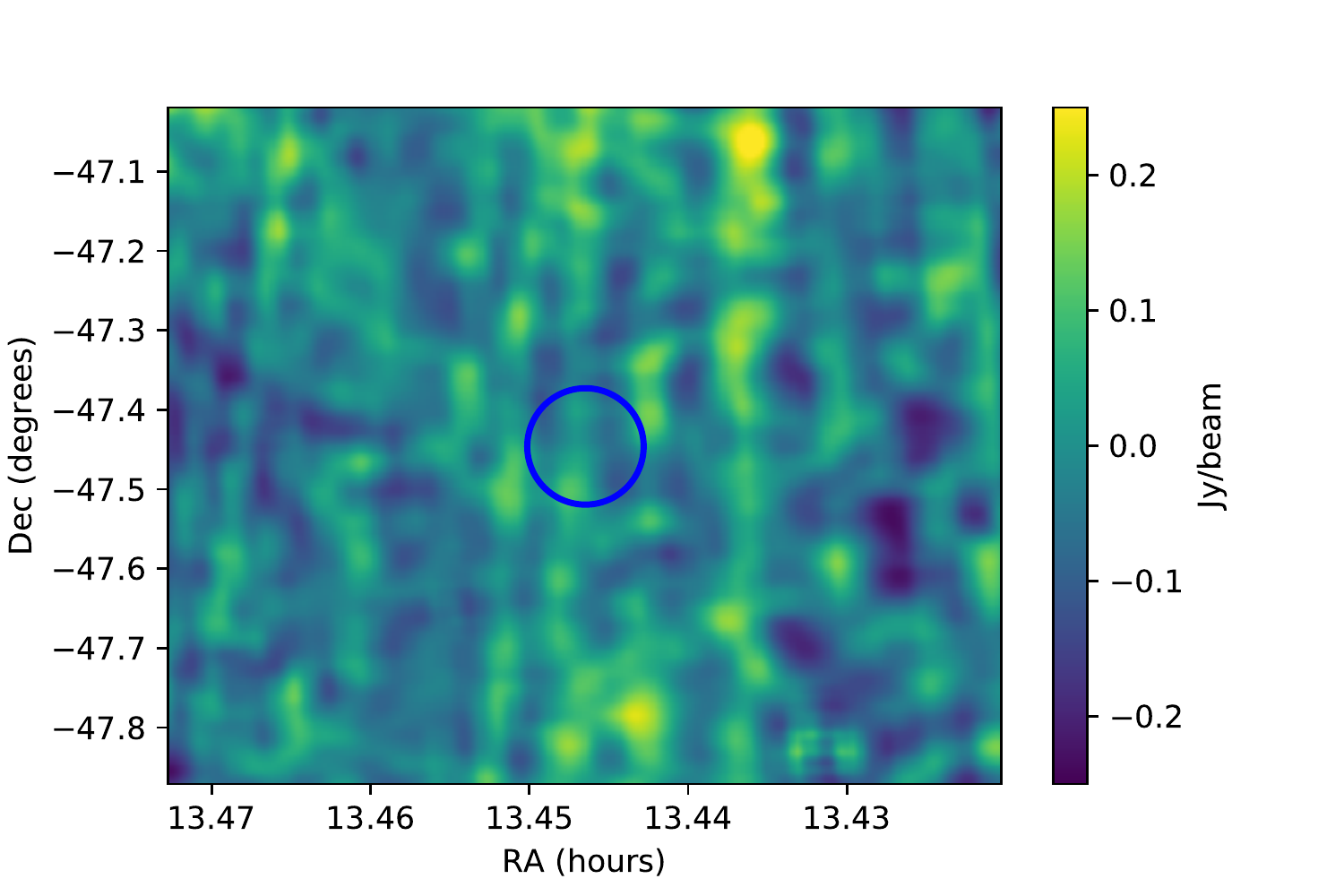}
\caption{Difference image for $\omega$ Cen using MWA and TGSS ADR1 data.  The blue circle denotes the optical half-light diameter for $\omega$ Cen.}
\label{radio-difference}  
\end{figure}

The difference image resulting from our analysis of the MWA data from Ref. \cite{2018MNRAS.474.4056M} and the GMRT TGSS ADR1 data of \cite{2017A&A...598A..78I} is shown in Figure \ref{radio-difference}.

While the images of the Cen A region by \cite{2018MNRAS.474.4056M} are of a very high quality, imaging artefacts at a low level are still apparent in Figure \ref{radio-difference}. These artefacts are in the form of radial stripes that originate near the peak intensity regions of Cen A. At the location of $\omega$ Cen, these stripes appear largely in a north-south orientation. The RMS of the pixel values in Figure \ref{radio-difference} is 72 mJy/beam, somewhat higher than the range of 14 - 63 mJy/beam achieved for 14 dSph galaxies by Ref. \cite{2019PhRvD.100d3002K}. Thus, we make some effort to remove the artefacts and lower the RMS.

Two approaches for artefact removal were employed. First, we applied a positional warp to the pixels of the image which varied as a function of pixel location, then applied a cubic interpolation function over the pixel brightness distribution at their new locations, effectively straightening the stripes in a north-south direction. We averaged across the pixel columns to calculate an average stripe profile, and subtracted this from every pixel row. Finally, the pixels were warped back to their original locations, and the image re-interpolated. Thus, with this approach some regions of the image are lost.  

The second approach calculates a generalized Hough transform of Figure \ref{radio-difference} that assumes the linear stripes converge at a single right ascension and declination coordinate pair. The transform finds the mean value of pixels along radial lines emerging from the assumed convergence point, producing a function of (average) surface brightness as a function of angle. The stripes are then reconstructed by interpolating the angular function at the location of each pixel, and the reconstructed stripes are then subtracted from Figure \ref{radio-difference}. 

\begin{figure*}[ht]
\centering
\begin{minipage}{0.48\textwidth}
  \includegraphics[height=0.6\textwidth,angle=0]{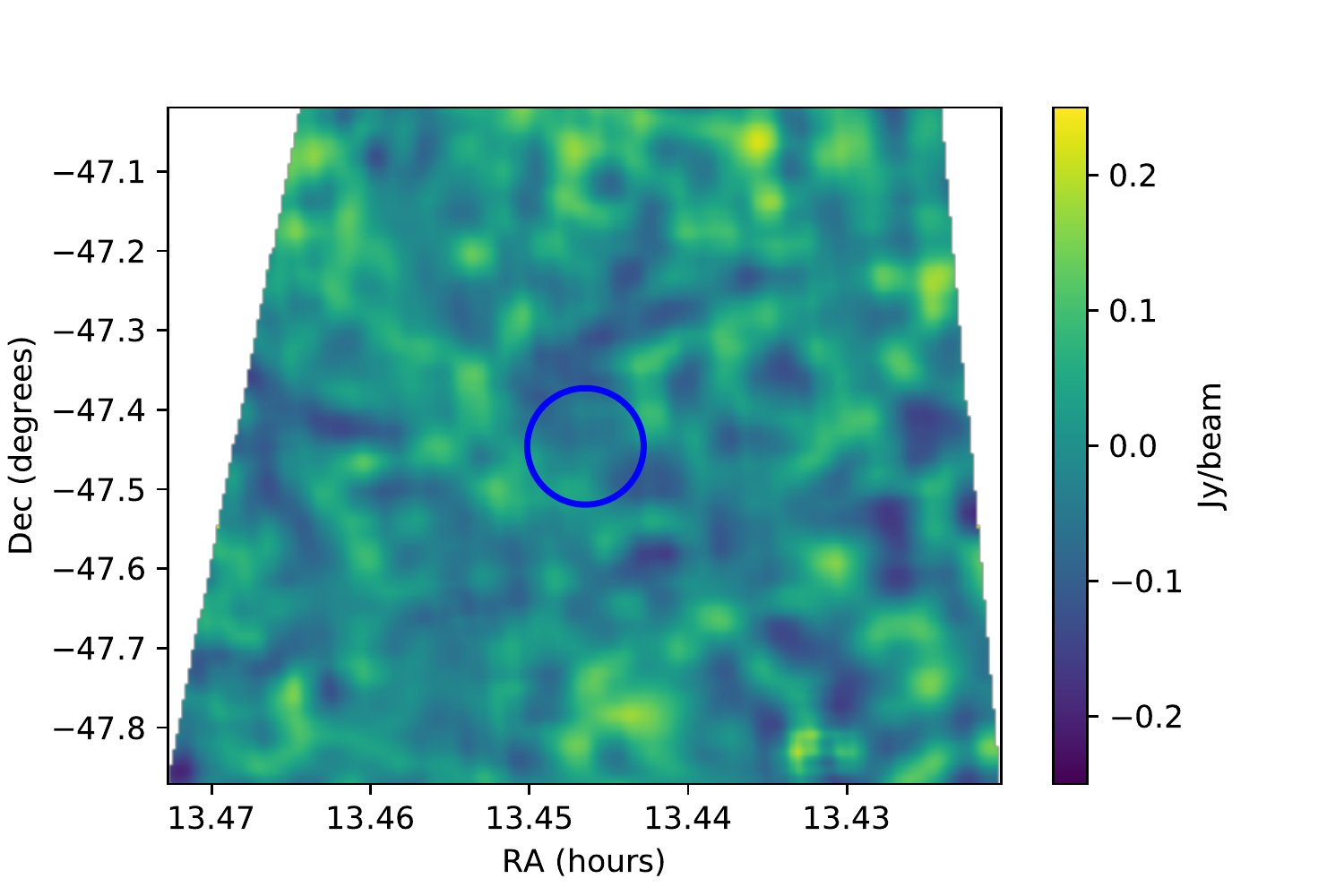}
\end{minipage}
\begin{minipage}{0.48\textwidth}
  \includegraphics[height=0.6\textwidth,angle=0]{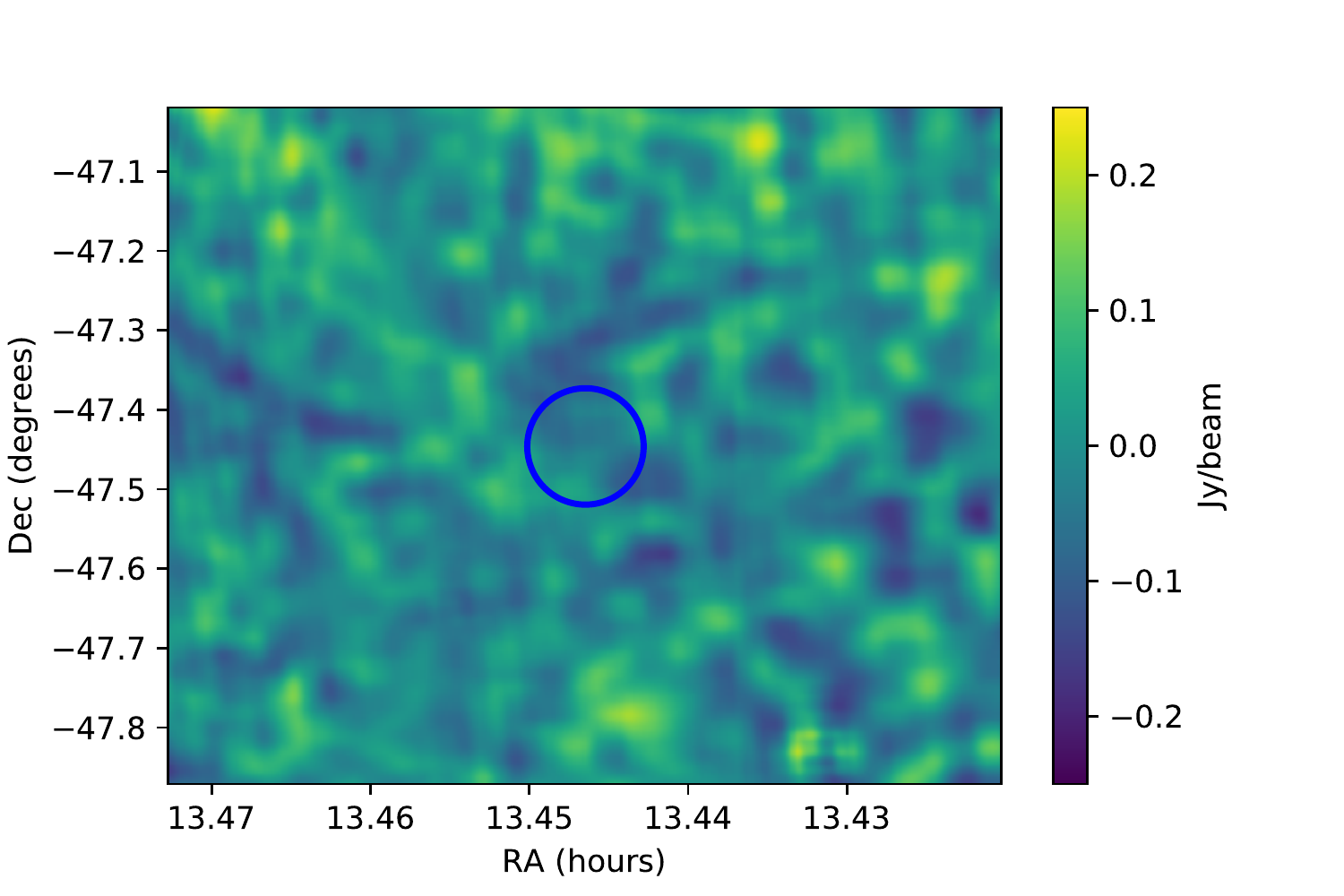}
\end{minipage}
\caption{Difference images for $\omega$ Cen following estimation and removal of artefacts.
The left panel shows the warp, average, and de-warp approach described in the text.
The right panel shows the residuals for the Hough transform approach described in the text.    The blue circles again denote the optical half-light diameter for $\omega$ Cen.}
\label{no-artefact}  
\end{figure*}

Both residuals are shown in Figure \ref{no-artefact}, where an excellent correspondence between the two approaches is evident, each producing a residual RMS of 58 mJy/beam, representing a 25\% improvement over the image in Figure \ref{radio-difference} and bringing the $\omega$ Cen observational limits within the range achieved by Ref. \cite{2019PhRvD.100d3002K}. At these limits, no evidence of diffuse synchrotron emission associated with $\omega$ Cen is evident.

\section{Radio emission model calculation}

DM annihilations inside $\omega$ Cen produce various standard model (SM) particles pairs (like $b\bar{b}$, $\mu^+ \mu^-$, $q\bar{q}$ etc.) which give rise to electrons-positrons ($e^{\pm}$) in their cascade decays. 
The abundance of these $e^{\pm}$ can be quantified by the source function \cite{Kar:2019cqo, PhysRevD.99.021302, McDaniel:2017ppt}
\begin{equation}
Q_e(E,r) = \frac{\langle \sigma v \rangle}{2 m_{DM}^2} \rho^2(r) \sum_{f} {\frac{dN^e_f}{dE}} B_f, 
\label{source_function}
\end{equation}
where $\rho(r)$ is the DM density profile at a radius $r$ from the centre of $\omega$ Cen. 
$\frac{dN^e_f}{dE} B_f$ determines the energy distribution of the $e^{\pm}$ produced per annihilation in the SM final state $f$ with branching fraction $B_f$.

The electrons(positrons), after being produced, diffuse through the interstellar medium and lose energy through various electromagnetic processes such as 
inverse Compton scattering, synchrotron effect, Coulomb effect, bremsstrahlung etc. The final equilibrium $e^{\pm}$ distribution ($\frac{dn_e}{dE} (E,r)$) is obtained by solving the 
transport equation \cite{Kar:2019cqo, McDaniel:2017ppt, 2007PhRvD..75b3513C, Colafrancesco:2005ji} 
\begin{equation}
D(E) \nabla^2 \left(\frac{dn_e}{dE}\right) +
\frac{\partial}{\partial E}\left(b(E) \frac{dn_e}{dE}\right) +
Q_e = 0.
\label{transeport_equation}
\end{equation}
Here $D(E)$ is the diffusion term and can be parameterised as 
$D(E) = D_0 \hspace{1mm} (\frac{E}{\rm GeV})^{0.3}$, where $D_0$ is the diffusion coefficient \cite{Kar:2019cqo, PhysRevD.99.021302, McDaniel:2017ppt, 2007PhRvD..75b3513C, 2019JCAP...03..019B}. 
The parameter $b(E)$ is the energy loss term which takes into account all the aforementioned energy loss processes (see references \cite{Kar:2019cqo, McDaniel:2017ppt, Colafrancesco:2005ji} for details). Note that, as the synchrotron loss 
increases with the magnetic field $B$, it causes the electrons-positrons to lose more energy for higher $B$ \cite{Kar:2019cqo}. 
Equation \ref{transeport_equation} can be solved by the Green's function method. Readers are referred to \cite{Kar:2019cqo, PhysRevD.99.021302, McDaniel:2017ppt, 2007PhRvD..75b3513C, Colafrancesco:2005ji} for the analytic form of 
the Green's function. 
The size of the diffusion zone for $\omega$ Cen is assumed to be $\sim 0.4$ kpc,
obtained by scaling with respect to the Segue I dSph \cite{McDaniel:2017ppt, 2015arXiv150703589N}.

Finally, the synchrotron surface brightness distribution 
$I_{\nu} (\theta)$ (at a frequency $\nu$), generated by the $e^{\pm}$ spectrum, is obtained by folding 
$\frac{dn_e}{dE} (E,r(l,\theta))$ with the synchrotron power ($P_{synch}(\nu,E,B)$) \cite{McDaniel:2017ppt, 2007PhRvD..75b3513C, Colafrancesco:2005ji, Natarajan:2013dsa} and integrating over 
the line-of-sight (los) \cite{2007PhRvD..75b3513C, Natarajan:2013dsa},
\begin{equation}
I_{\nu}(\theta)=\frac{1}{4\pi}\int_{\rm los}dl\left(2\overset{m_{DM}}{\underset{m_e}{\int}}dE\frac{dn_e}{dE} P_{synch}\right),
\label{surface_brightness}
\end{equation}
where $l$ is the line-of-sight coordinate. As described in \citep{2019PhRvD.100d3002K}, this surface brightness distribution is then convolved with the Phase I MWA beam and the 
peak value is compared with the observation.

One of the primary ingredients in our calculation is a DM profile for $\omega$ Cen 
(see equation \ref{source_function}). In this analysis, we have assumed a Navarro-Frenk-White (NFW) profile ($\rho (r)$) \cite{Navarro:1995iw} for the DM distribution associated with $\omega$ Cen,

\begin{equation}
  \rho (r) = \rho_s \left(\frac{r}{r_s}\right)^{-1} \left(1 + \frac{r}{r_s}\right)^{-2}  ,
 \label{rho_profile} 
\end{equation}
with $r_s$ and $\rho_s$ being the scale radius and density, respectively. The astrophysical J-factor, which determines the number of DM pairs available to annihilate inside $\omega$ Cen, is defined as the line-of-sight (los) integration 
of $\rho^2 (r)$,

\begin{equation}
  \mbox{J} = \int_{\Delta \Omega} d\Omega \int_{\rm{los}} \rho^2 (r (l, \Omega)) dl  ,
\label{Jfactor}  
\end{equation}
where $\Delta \Omega$ is
the angular size for the DM distribution in $\omega$ Cen \cite{2019arXiv190708564B}.
Using stellar kinematics data for $\omega$ Cen, Ref. \cite{2019arXiv190708564B} has obtained $r_s$ and the J-factor within 68\% confidence limits.

The implications of the claims made by Ref. \cite{2019arXiv190708564B} and Ref. \cite{Reynoso-Cordova:2019biv}
for MWA data can be best assessed if one uses the same J-factor to study both. We
thus  use the best-fit J-factor (assuming a NFW profile) obtained from  stellar kinematics in Ref. \cite{2019arXiv190708564B}, via a method independent of radio or $\gamma$-ray data. 
This value is used also in the case of  
Ref. \cite{Reynoso-Cordova:2019biv}, where the best-fit value of $\langle \sigma v \rangle \rm{J}$ is listed.
The best-fit value of $\langle \sigma v \rangle$ is thus extracted, assuming a diagonal 
correlation matrix. The corresponding
value is already extracted in  the analysis of Ref. \cite{2019arXiv190708564B}. These quantities,
together with the DM mass $m_{DM}$ and the dominant annihilation channel, comprise
the model inputs.

In our calculations for the expected synchrotron signal from DM annihilation, we have used three different sets of $r_s$ and $\rho_s$ which are denoted as [$r_{s_0}$,$\rho_{s_0}$], [$r_{s_{max}}$,$\rho_{s_{min}}$], and [$r_{s_{min}}$,$\rho_{s_{max}}$]. 
Here, $r_{s_0}$, $r_{s_{max}}$, and $r_{s_{min}}$ are the best-fit and 68\% maximum and minimum values of $r_s$ from Ref. \cite{2019arXiv190708564B}, respectively
\footnote{$\lbrace r_{s_0}, r_{s_{max}}, r_{s_{min}}\rbrace$ $=$ $\lbrace 1.63, 4.63, 0.11\rbrace$ (in $\rm{pc}$); \\
$\lbrace \rho_{s_0}, \rho_{s_{min}}, \rho_{s_{max}}\rbrace$ $=$ 
$\lbrace 2.87\times10^5, 6.03\times10^4, 1.64\times10^7\rbrace$ (in $\rm{GeV cm^{-3}}$).}. They produce the same best-fit J-factor found in the analysis of 
Ref. \cite{2019arXiv190708564B}. One can verify numerically that to a close approximation the J-factor scales as $\rho^2_s r^3_s$.  At the same time, the Green's function, relevant to the calculation
of the peak radio surface brightness, scales in the same manner. 
Thus, once one adopts the best-fit value of the J-factor,
and changes $r_s$ with $\rho_s$ altered concomitantly to keep J the same, one should obtain
the same peak surface brightness, as can be seen in Table \ref{Table_2}, using illustrative values of the diffusion coefficient $D_0$ and the magnetic field $B$. 
The results presented henceforth use the best-fit values of the NFW parameters, but are valid over the 68\% confidence interval.


\begin{table*}[ht!]
\begin{center}
\def\arraystretch{1.5}%
\def\columstretch{1.5}%
\begin{tabular}{|c|c|c|c|c|c|}
\hline
$D_0$ & $B$ & channel & I (mJy/beam) & I (mJy/beam) & I (mJy/beam) \\
 ($\mbox{cm}^2 \mbox{s}^{-1}$) & ($\mu G$) &  & with [$r_{s_0}$,$\rho_{s_0}$] & with [$r_{s_{max}}$,$\rho_{s_{min}}$] & with [$r_{s_{min}}$,$\rho_{s_{max}}$] \\
\hline

$3 \times 10^{26}$ & 5 & $b\bar{b}$ & 15.1 & 14.9 & 15.2 \\
& & $q\bar{q}$ & 9.8 & 9.7 & 9.7 \\
& & $\mu^+\mu^-$ & 1962.4 & 1922.3 & 1979.6 \\
\hline

$3 \times 10^{27}$ & 10 & $b\bar{b}$ & 4.6 & 4.6 & 4.7 \\
& & $q\bar{q}$ & 3.9 & 3.9 & 4.0 \\
& & $\mu^+\mu^-$ & 480.4 & 475.1 & 482.6 \\
\hline
\end{tabular}
\caption{Columns 1 and 2: Some illustrative values of diffusion coefficient ($D_0$) and magnetic field ($B$) in $\omega$ Cen; 
Column 3: Different best-fit annihilation channels from \cite{2019arXiv190708564B} 
($b\bar{b}$)
and \cite{Reynoso-Cordova:2019biv} ($q\bar{q}$ and $\mu^+\mu^-$);
Columns 4 - 6: Predicted peak synchrotron surface brightnesses (I (mJy/beam)), convolved with the MWA beam,
corresponding to those values of $D_0$ and $B$ (mentioned in columns 1 and 2) for different choices of [$r_s$,$\rho_s$] which produce the same best-fit J-factor
for $\omega$ Cen as quoted in \cite{2019arXiv190708564B}. 
The DM mass ($m_{DM}$) and annihilation rate ($\langle \sigma v \rangle$)
in each case have been fixed at the values obtained from Ref. \cite{2019arXiv190708564B} and Ref. \cite{Reynoso-Cordova:2019biv}.}
\label{Table_2}
\end{center}
\end{table*}

\section{Magnetic field estimate for $\omega$ Cen}

As noted earlier, an important element in modeling the synchrotron emission due to DM annihilation is the ambient magnetic field strength.
In the case of our previous work on DM annihilation in dSph galaxies, producing estimates of the magnetic field strength was difficult.
In the current work, because $\omega$ Cen belongs to our Galaxy, we can use the best available models of the Galactic magnetic field to
make independent estimates, thereby constraining our DM annihilation models.

We estimate the interstellar magnetic field at the location of $\omega$ Cen using two leading Galactic magnetic field models: the model
by Jansson and Farrar \cite{2012ApJ...757...14J, 2012ApJ...761L..11J} (hereafter JF12) and the model
by Jaffe {\it et al.} \cite{2013MNRAS.431..683J} (hereafter J13).
These are two of the most advanced models of the magnetic field in the Milky Way. They both consist of a regular magnetic field component,
an isotropic turbulent component, and an anisotropic turbulent component which has a random direction but a fixed orientation along the
regular magnetic field. The two models are in many ways similar, but each have their strengths and weaknesses. 

The regular magnetic field is assumed to have a spiral shape in both models, with slightly different parametrizations.
The magnetic field runs along the spiral arm and its strength is allowed to vary between segments. The two approaches make
partially different choices of data to fit the models to, resulting in different uncertainties (see Ref. \cite{2014A&A...566A..55P}
for a detailed discussion). JF12 includes also an out-of-plane component of the regular magnetic field, which is ubiquitously
observed in edge-on nearby spiral galaxies as the so-called X-shaped field. J13 does not have an out-of-plane component,
but fits also to dust polarization, allowing differentiation between locations of the synchrotron arms and the dust arms.
For both models, we use the updated best-fit parameter values which match the Planck synchrotron data \cite{2014A&A...566A..55P}.

\begin{figure}[t!]
\centering
  \includegraphics[height=0.3\textwidth, angle=0]{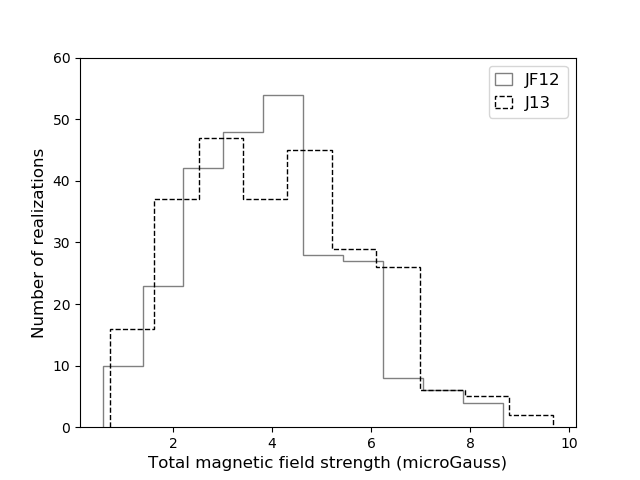}
\caption{Histograms of the total magnetic field strength at the location of $\omega$ Cen, in 250 realizations of the random field component, for the Galactic magnetic field models of JF12 and J13}
\label{f:bfieldmodels}
\end{figure}

For both models, we use the {\sc Hammurabi} software package \cite{2009A&A...495..697W} to calculate the Galactic magnetic
field at the location of $\omega$ Cen (assumed to be at $(\ell,b) = (309^{\circ}, 15^{\circ})$ and at a distance of 5~kpc).
We use the best-fit parameters of these models to calculate the regular magnetic field component, which is consistent between
the two models at a value of $B_{reg}\approx 1~\mu$ G at this location. For both models, we ran 250 realizations of the random
magnetic field component to gauge the range of possible magnetic field values at the location of $\omega$ Cen. These distributions
are shown in Figure~\ref{f:bfieldmodels}.


We adopt magnetic fields between 1 and 10 $\mu$G as a plausible range over which to calculate our DM annihilation models.

\section{Results}

\begin{figure*}[ht!]
\centering
  \includegraphics[height=0.3\textwidth, angle=0]{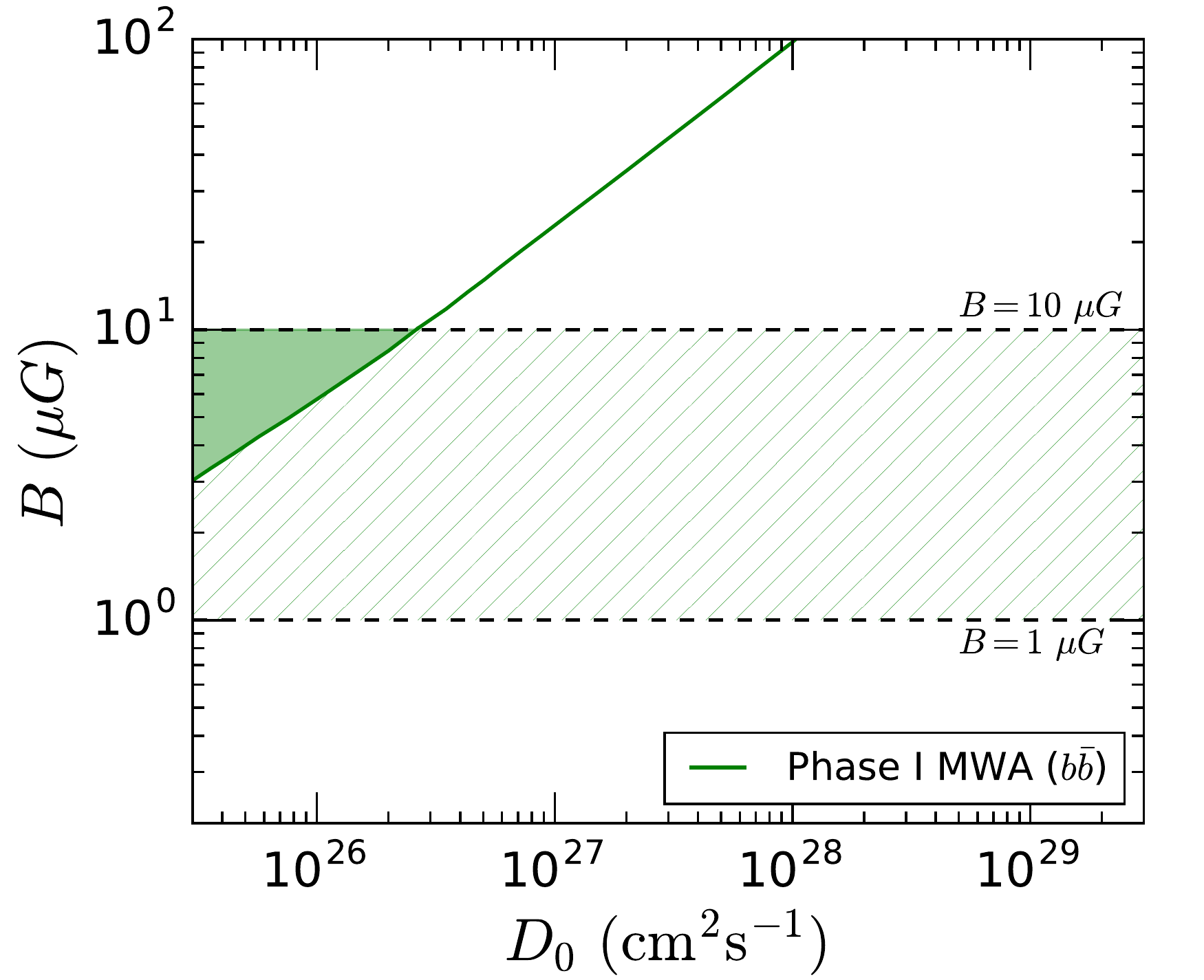}\hspace{8mm}
  \includegraphics[height=0.3\textwidth, angle=0]{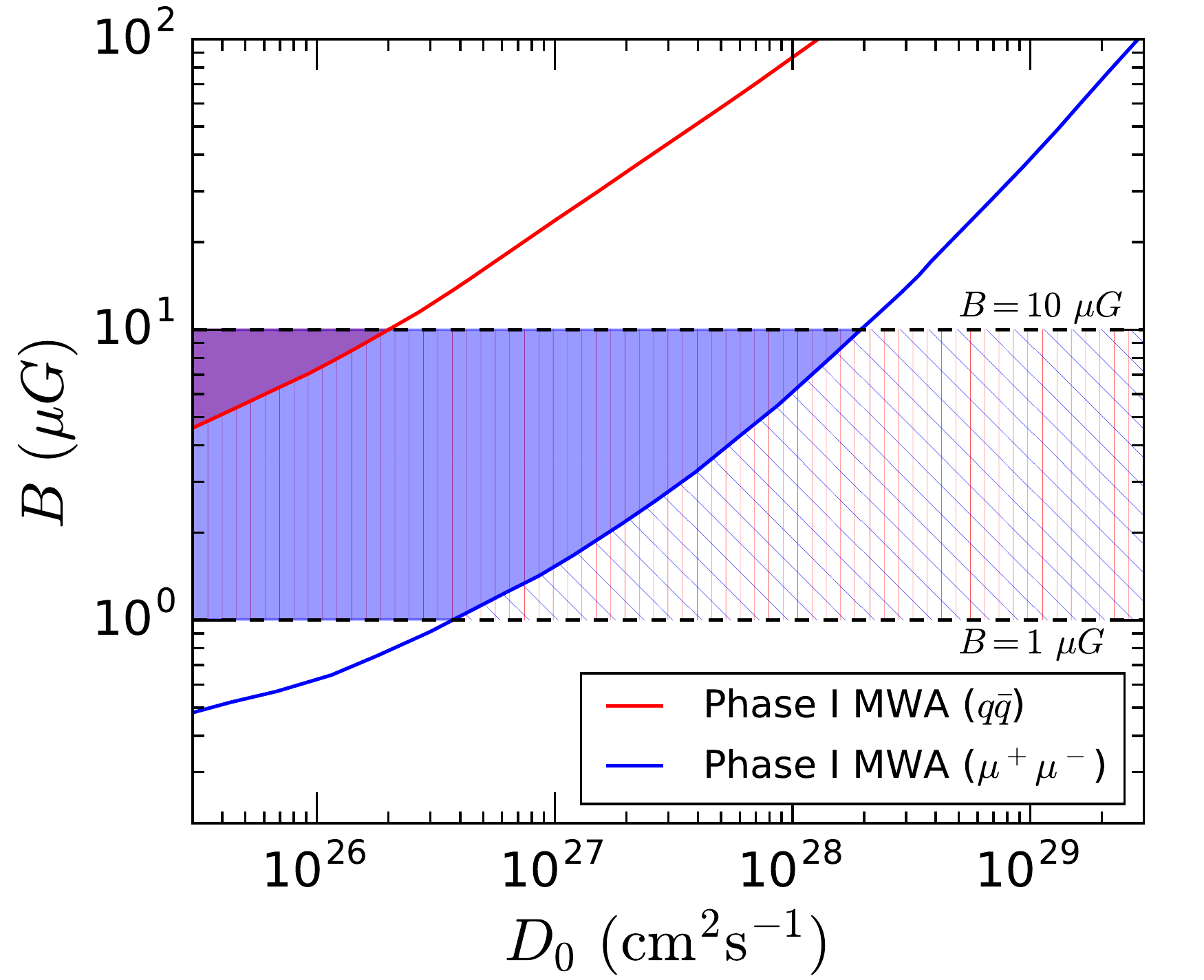}
\caption{Areas in the $B - D_0$ plane (denoted by shaded colors) which are ruled out by the Phase I MWA data on $\omega$ Cen. {\it Left panel:} Corresponds to the findings of \cite{2019arXiv190708564B} where the DM annihilates dominantly into $b\bar{b}$. {\it Right panel:} Shows the corresponding regions, based on \cite{Reynoso-Cordova:2019biv}, for DM annihilation into $q\bar{q}$ (red) and $\mu^+\mu^-$ (blue). 
The range of $B$ is taken as 1 - 10 $\mu G$ (see the foregoing analysis).
In each case, the hatched area between 1 - 10 $\mu G$ indicates the region which is still allowed by the MWA I data.}
\label{B_D0_limit}  
\end{figure*}

Figure \ref{B_D0_limit} shows the constraints on the $B - D_0$ plane for $\omega$ Cen, obtained
from our MWA data. The shaded areas correspond to cases where the predicted peak surface brightness is higher than 58 mJy/beam, the observational limit. As estimated earlier, the upper limit on the magnetic field ($B$) at $\omega$ Cen is taken to be 10 $\mu G$ and the lower limit is taken to be 1 $\mu G$ (shown by the horizontal dashed 
lines). The $\mu^\pm$ annihilation scenario of \cite{Reynoso-Cordova:2019biv} is tightly constrained, which is relaxed for the $q\bar{q}$ channel. The
$b\bar{b}$ channel suggested in \cite{2019arXiv190708564B} falls in between, having a slightly
larger exclusion region than in the second case above.

Figure \ref{B_D0_limit} is instructive since it places limits on relatively low values of $D_0$ in $\omega$ Cen, whose vicinity to the Milky Way would make such values otherwise plausible, in contrast to dSph galaxies \cite{Natarajan:2013dsa,Regis:2014koa}. We can also use it to explore the effect of deeper observational limits on the synchrotron emission. For example, if a limit ten times deeper (5 mJy/beam) could be achieved, values of $D_{0}$ $\simeq 10$ times larger could be ruled out for $B = 10 \mu G$. We note that our early processing of data using the upgraded MWA \cite{2018PASA...35...33W} at the position of $\omega$ Cen indicates that improvements of this order appear likely, which will be explored in a future publication.  A limit 100 times deeper (0.5 mJy/beam) would strengthen the limit correspondingly, bringing the models into serious conflict with observations. In this case, such improvements will only likely be made using the future capabilities of the SKA. Improved knowledge of the magnetic field would assist us even further in this respect.

Using the best-fit values of $m_{DM}$ and $\langle \sigma v \rangle$ we can briefly revisit the analysis of the Boo dSph galaxy from our previous work \cite{2019PhRvD.100d3002K}.  
Note that, among the 14 dSph studied previously, we obtained largest predicted radio signal for Boo.
In the current work, with the best-fit DM annihilation model corresponding to the $\gamma$-ray observation, we find that the predicted synchrotron signal for Boo still lies below the threshold for detection with the MWA we reported.  The parameters that produce a prediction closest to our observational limits are ($B\gtrsim 10 \mu G, D_0 \lesssim
10^{26} \mbox{cm}^2 \mbox{s}^{-1}$), for the model corresponding to annihilation into $\mu^+\mu^-$.


\section{Conclusion}

We conclude that, following the studies in \cite{2019arXiv190708564B} and \cite{Reynoso-Cordova:2019biv}, objects such as $\omega$ Cen are likely to provide some of the best environments in which to test DM annihilation models with $\gamma$-ray and radio synchrotron observations
in tandem, plus studies of magnetic field strength. $\omega$ Cen produces a much stronger predicted synchrotron signal than typical dSph galaxies, for the same annihilation models, and the prospects for understanding the magnetic field strength in galactic objects such as $\omega$ Cen is better. 
We therefore derive significant limits on model parameters from current MWA data.
Also, observations deeper than those used here can subject DM annihilation models to serious tests.

The results presented here provide one avenue among a rich and growing set of investigations into the presence and nature of DM, now including multi-wavelength and multi-messenger approaches \citep{2020PhLB..80535439A}.  
In general, a diverse set of empirical approaches is required, as DM properties are unclear and a wide span of theoretical model parameter space confronts observations.  
As such, exploratory studies of electromagnetic and non-electromagnetic signatures of DM processes are required to match this theoretical uncertainty.

\section*{Acknowledgements}
AK and BM were partially supported by funding available from the Department of Atomic Energy, Government of India, for the Regional Centre for
Accelerator-based Particle Physics (RECAPP), Harish-Chandra Research Institute.
MH acknowledges funding from the European Research Council (ERC) under the European Union’s Horizon 2020 research and innovation programme (grant agreement No 772663).


\section*{References}


\end{document}